# A Perspective on Recent Advances in Phosphorene Functionalization and its Application in Devices

Maurizio Peruzzini,*[a] Roberto Bini,[a,b,c] Margherita Bolognesi,[d] Maria Caporali,[a] Matteo Ceppatelli,[a,b] Francesca Cicogna,[e] Serena Coiai,[e] Stefan Heun,[f] Andrea Ienco,[a] Inigo Iglesias Benito,[a,g] Abhishek Kumar,[f] Gabriele Manca,[a] Elisa Passaglia,[e] Demetrio Scelta,[a,b] Manuel Serrano-Ruiz,[a] Francesca Telesio,[f] Stefano Toffanin[d] and Matteo Vanni[a,g]

**Abstract:** Phosphorene, the 2D material derived from black phosphorus, has recently attracted a lot of interest for its properties, suitable for applications in material science. In particular, the physical features and the prominent chemical reactivity on its surface render this nanolayered substrate particularly promising for electrical and optoelectronic applications. In addition, being a new potential ligand for metals, it opens the way for a new role of the inorganic chemistry in the 2D world, with special reference to the field of catalysis. The aim of this review is to summarize the state of the art in this subject and to present our most recent results in preparation, functionalization and use of phosphorene and its decorated derivatives. In particular, we discuss several key points, which are currently under investigation: the synthesis, the characterization by theoretical calculations, the high pressure behaviour of black phosphorus, as well as decoration with nanoparticles and encapsulation in polymers. Finally, device fabrication and electrical transport measurements are overviewed on the basis of recent literature and new results collected in our laboratories.

## 1. Introduction

Discovered for the first time by the alchemist Brandt in 1669,[1] phosphorus is the 11$^{th}$ most abundant element on the Earth's crust, with a total concentration of 1120 ppm.[2] Due to its high reactivity with oxygen, no elemental form of phosphorus is available in nature in contrast with other main-group elements, such as carbon[3] or sulphur.[4] In nature, phosphorus is always available as phosphates, whose importance overcomes the frontiers of pure inorganic chemistry. Phosphates and their derivatives are central in many industrial sectors, such as cleaners, flame retardants and, in particular, fertilizers.

Next to the industrial relevance, it is noteworthy to point out that phosphates are ubiquitous in all living beings, in the energy exchange chains, as ATP, in the DNA or RNA building blocks, and in many essential processes such as bones and teeth formation and growth.[5]

The technique used worldwide to obtain elemental phosphorus from mineral phosphorites is the high temperature treatment of phosphates with suitable reducing agents, mainly carbon coke. The final product is white phosphorus, which consists of discrete tetrahedral $P_4$ molecules, Figure 1, with an associated lone pair on each P atom. White phosphorus is a particular added-value product, being the precursor of the industrially relevant organophosphorus compounds, after treatment with molecular chlorine followed by reactions with suitable organic substrates for the introduction of organic substituents at P centres.[6]

By heating white phosphorus, the red allotrope is achieved, which shows a polymeric disordered structure after some P-P bond breaking of the parent $P_4$ moiety followed by formation of new P-P links among the different units. Further heating and treatment with suitable catalytic reagents, as iodine or $I_2$ sources (*vide infra*), transforms red phosphorus into a third phosphorus allotrope, namely black phosphorus, bP.[7]

A schematic chart of some interconversions among the different phosphorus allotropes is depicted in Figure 1.

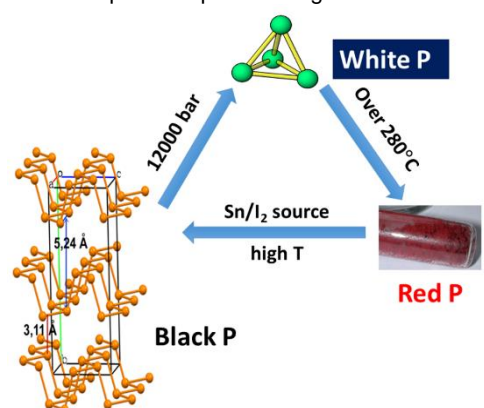

Figure 1. Chart of modification of phosphorus allotropes varying temperature T and pressure P.

Black phosphorus shares with red phosphorus a high degree of polymerization, although, in contrast with the red allotrope, it shows an ordered layered structure, with inter-layer distances of 3.11 Å.[8] Each layer shows a puckered structure with three-coordinated phosphorus atoms and a lone pair available for each P atom. For many decades after its discovery, bP has been marginally investigated as shown in Figure 2, and only around 2015 we assisted to a blooming attention by chemists and physicists, which has constantly increased in the following years (see Figure 2).

[a]  Consiglio Nazionale delle Ricerche – Istituto di Chimica dei Composti Organometallici
Via Madonna del Piano 10, 50019 Sesto Fiorentino, Florence, Italy
E-mail: maurizio.peruzzini@cnr.it
Web: http://www.dsctm.cnr.it/en/department/director.html
[b]  LENS- European Laboratory for Non-Linear Spectroscopy, Via N. Carrara 1, I-50019, Sesto Fiorentino (FI), Italy.
[c]  Dipartimento di Chimica "Ugo Schiff, Università degli Studi di Firenze, Via della Lastruccia 3, I-50019, Sesto Fiorentino (FI), Italy.
[d]  Consiglio Nazionale delle Ricerche – Istituto per lo Studio dei Materiali Nanostrutturati
Via Piero Gobetti, 101, 40129 Bologna BO, Italy
[e]  Consiglio Nazionale delle Ricerche – Istituto di Chimica dei Composti Organometallici, SS Pisa
Via Moruzzi 1, 56124 Pisa, Italy
[f]  NEST, Istituto Nanoscienze-CNR and Scuola Normale Superiore, Piazza San Silvestro 12, 56127 Pisa, Italy
[g]  Dipartimento di Biotecnologie, Chimica e Farmacia, Università di Siena, via A. Moro, 2, 53100 Siena, Italy

Such a bP "renaissance" in 2014-2015 was mainly due to the discovery of the astounding electronic and physical properties of the exfoliated mono-layer of bP, namely phosphorene, (blue bars in Figure 2), which is the focus of this review.[9-11]

In recent years, 2D materials have become a very hot research topic in science after the seminal discovery of graphene, the monolayer material formed by carbon atoms obtained by graphite exfoliation.[12] Although particularly interesting in view of the high thermal conductivity and carriers mobility, graphene suffers from the lack of an electronic bandgap, achieved only through breaking of crystal symmetry, *e.g.* via hydrogenation or application of an electric field.[13] Research on graphene triggered the discovery of new 2D materials, and, among them, exfoliated bP seems to be particularly appealing in view of its bandgap, the value of which depends on the number of stacked layers. In this regard, the direct bandgap may be tuned from 0.3 eV for bulk bP up to *ca.* 2 eV for the mono-layer, phosphorene.[14]

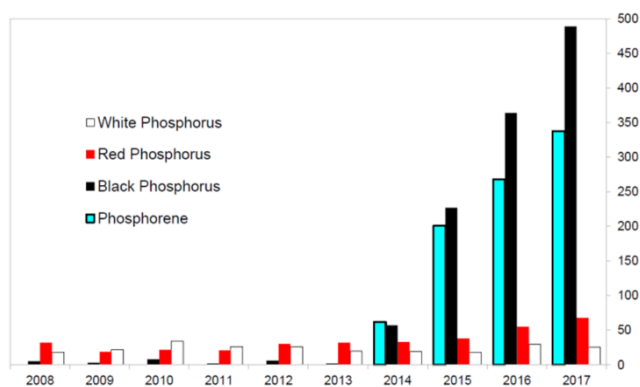

Figure 2. Distribution of research articles associated to the different phosphorus allotropes in the 2008-2017 decade. Source: Scopus.

The main drawback of exfoliated black phosphorus is its high reactivity, particularly when exposed to air, which easily results in the formation of phosphates.[14] In recent years, large efforts have been devoted to the search of efficient strategies of protective modifications of the exfoliated bP in order to enhance its stability and to develop electronic and optical devices based on it.

Since some decades there is in Florence a strong research tradition on phosphorus chemistry, mainly addressing the reactivity of $P_4$ and the development of new synthetic strategies to obtain valuable P-containing products via sustainable processes. Based on the know-how acquired in the last years, at the end of 2014, just few months before the blooming of the articles about the bP and phosphorene, our group applied for and won a European Research Council (ERC) Advanced Grant aimed at investigating the functionalization of the exfoliated black phosphorus. The project, entitled *"Phosphorene functionalization: a new platform for advanced multifunctional materials"*, nicknamed PHOSFUN, had the main aim to explore the chemical reactivity of bP and phosphorene towards inorganic and organic reagents, as well as to search for its potential applications. The main strength of the project is that it combines chemistry and physics know-how, allowing a complete panorama of all aspects of bP and its exfoliated form. This review intends to provide a summary of the research activity carried out on phosphorene with particular emphasis to our PHOSFUN project.

Alike many other groups trying to crack the so far hard nut of the chemical functionalization of phosphorene, we also have faced and are facing important difficulties, mostly due to the high instability of this material towards light, oxygen and water. However, it is now well acquired that there are many possibilities to stabilize exfoliated bP and even phosphorene, which are paving the way to their use for electronic and optical applications while unexpected biomedical applications are also emerging.

## 2. Synthesis of Black Phosphorus and Phosphorene

Black phosphorus (bP) was serendipitously discovered in 1914 by Bridgman[15] while attempting to synthesize red phosphorus starting from the white allotrope using pressures between 1.2 GPa - 1.3 GPa at 200 °C. Alternative routes were also explored using both white and red phosphorus at pressures between 4 - 4.5 GPa and temperatures between 400 – 800 °C [16] up to the use of only red-P at a pressure of 1 GPa between 900 – 1000 °C.[9,17] The electrical properties of black phosphorus depend greatly on its purity, degree of crystallinity and defects-vacancy, and these in turn depend not only on the synthesis method, but also on the starting reagents.[18] The first synthesis at low pressure was reported in the 1960s and later improved in the 1980s.[19] bP was obtained by recrystallization of white-P in liquid Bismuth at 300-400°C. Unfortunately, the crystals were needles (5 x 0.1 x 0.01 mm$^3$) smaller than those obtained by high-pressure routes. In 2007, Nilges *et al.* found that bP could be synthesized from red-P by the Chemical Vapor Transport mechanism at low pressure and temperatures between 600°C – 650°C.[7a] This was a great leap not only for the quality of the material obtained (size and crystallinity), but also for the simplicity of the synthesis. Just one year after,[7b] an improvement of the method was published. In 2014, the synthesis using only Sn and $SnI_4$ as mineralizer was published.[20] Basically red-P, a small quantity of a tin-gold alloy or tin and $SnI_4$ were placed inside a quartz tube sealed under inert atmosphere at low pressure. The sealed ampoule was heated to 400°C for 2h, then the ampoule was further heated up to 650 °C and maintained at this temperature for 70 hours. Afterwards the temperature was reduced to 500°C at rate of 40 °C/h, before being cooled down to room T within 4h. In our laboratory, we found that crystals with high crystallinity and purity were obtained using a slightly longer reaction time (72h) and a slower cool down (0.1 °C/min) (Figure 3).

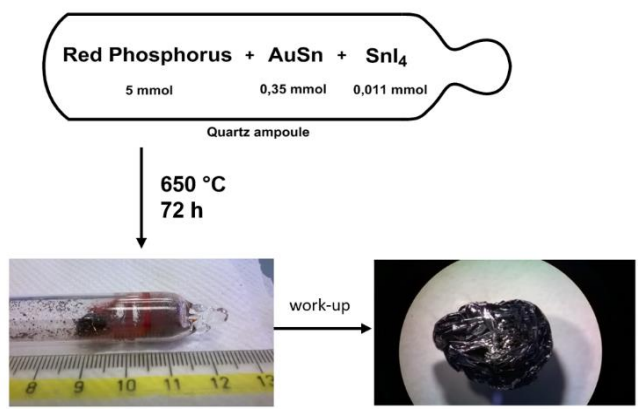

Figure 3. Schematics of the synthesis of black phosphorus

Recently large crystals of bP have been obtained through a ternary clathrate $Sn_{24}P_{22-x}I_8$ at 550°C and at low pressure in a sealed ampule.[21]

The great interest and importance, aroused by black phosphorus in the scientific community at the beginning of 2014, was due to the fact that two research groups, at almost the same time, for the first time reported the isolation of bP layers with thickness between 2 nm – 10 nm.[9,10] They found that the fascinating electronic properties of this material depend on the number of layers. The technique used, the scotch-tape based micro-cleavage of bulk bP (mechanical exfoliation), was the same as that employed for graphene in 2004 by Geim and Novoselov.[12] For large scale production, alternative developed methods for preparing phosphorene (or few layer bP, 2D bP) are: 1) liquid phase exfoliation (LPE), 2) shear assisted exfoliation, 3) plasma assisted exfoliation, 4) pulsed laser deposition, 5) microwave assisted exfoliation, 6) electrochemical exfoliation and 7) ultrasonication mediated by mechanical milling.[22] Nevertheless, mechanical exfoliation remains the most used synthetic approach for electronic applications, while for other applications (sensors, biology,[23] catalysis, photo-catalysis, energy storage and production, *etc*.), LPE is mainly employed.[24] LPE in fact provides a high yield of bP few-layers in solution with an efficient, low cost, and simple approach. The most used LPE-solvent is NMP (N-methylpyrrolidone), however other solvents like DMSO (dimethyl sulfoxide) give good quality exfoliated bP flakes.[25]

Recently it has been demonstrated that 2D bP nanoflakes can be directly prepared by a solvothermal reaction of white P in ethylenediamine, without any additives.[26] Using the same protocol, 2D bP nanoflakes decorated by cobalt phosphide nanoparticles have been also directly synthetized.[27] In principle this approach is making the fabrication facile, low-cost and easily scalable.

## 3. Theoretical studies

Nowadays, computational investigation has become a powerful tool in chemistry for both providing reasonable explanations of experimental observations and predicting new reaction pathways, to be later addressed and confirmed in the laboratory. In particular, in the case of 2D materials, computations are particularly useful to highlight electronic structure and the feasible reactivity with peculiar substrates in order to infer particular electronic and physical properties to the resulting material.[28] For the exfoliated bP, the possibility to tune the electronic structure by changing the number of the stacked layers as well as the high tendency to be oxidized by reacting with oxygen makes computational chemistry an intriguing tool for the investigation of the material.[29] In this regards, many computational efforts have been addressed toward the understanding of the electronic structure and all the possible variations of physical and electronic properties associated to external perturbations, such as the application of an electric field[9] or strain or doping with heteroelements like boron or carbon. For example, the computational investigation revealed how the bandgap is stepwise reduced by stacking more layers of phosphorus.[14] Based on *in-silico* results, the doping of the material with boron atoms in place of phosphorus ones has been demonstrated to reduce the bandgap and to allow the potential activation of $H_2$ molecules with the possible formation of frustrated Lewis-pairs useful for the catalytic hydrogenation of unsaturated substrates, such as ketones.[30]

Although the availability of a large amount of lone pairs on the surface should in principle suggest the possibility of functionalization, thus enhancing the stability of the material, such a feature was only scarcely addressed from both a computational and experimental viewpoint. Some examples are available but most of them concern the adsorption of delocalized organic onto the 2D surface[31] while for the covalent functionalization at the best of our knowledge only few studies are known. The available experimental examples concern the functionalization of the surface with diazonium salt with release of $N_2$ or with $TiL_4$ compounds. In both cases spectroscopic evidences are provided. [32] Two other available studies are related to the computational investigation of the interaction modes of simple metal atoms with the phosphorene surface and the implicit consequences on the band structure[33] and on the absorption of chromium trioxide.[34]

In view of the know-how acquired over the years on the phosphorus compounds,[35] we started an *in-silico* voyage within PHOSFUN project, aimed at highlighting the properties and the reactivity of phosphorene and its functionalization. The philosophy of such a computational analysis was to develop a detailed knowledge of the energy/properties relationships, taking into account the orientation of the lone pairs of the 2D material surface, being not perpendicular but bent by 30°. A potential way for the introduction of functional groups on the phosphorene surface should be the reaction between the exfoliated material with halide sources in order to create P-X bonds, which can be later further functionalized through suitable substitution reactions. In view of the well documented affinity of phosphorus toward iodine and on our know-how on the halogen bonding,[36] we wondered if the reaction between exfoliated black phosphorus and molecular iodine, as iodide source, could result in the material functionalization. In this regard, we continuously compared the behaviour of the single layer of phosphorene with other phosphorus compounds, containing P atoms in a local $sp^3$ pyramidal connection with three other P atoms, such as white phosphorus or phosphines ($PR_3$). A strong base such as a $PR_3$ interacts with $I_2$ forcing a prompt charge separation and the formation of an ion pair of formula $[PR_3I^+][I^-]$ with an estimated free energy gain as large as -20 kcal mol$^{-1}$.[36] Additional gain may be obtained by the involvement of a second $I_2$ molecule, being the driving force the consequent generation of the triiodide anion. For $P_4$, all the textbooks report the high energy gain and the completeness of the reaction with molecular iodine to provide $PI_3$, although the mechanism has never been elucidated in details. We were able to depict the complete flowchart of the $P_4$ transformation to $PI_3$, through the isolation of all possible intermediates. The process consists in a sequential P-P activation with a concerted mechanism, involving two $I_2$ molecules with the cleavage of a P-P and two P-I bonds and the formation of two P-I and one I-I bond. The only detected energy barrier is for the first P-P cleavage, after that the process is a cascade. We applied a similar approach to the phosphorene layer..[37] The analysis revealed the reasonable formation of a weak adduct between one P lone pair of the material and an $I_2$ molecule, although the P---I distance remains rather long (3.2 Å), suggesting a somewhat feeble donor power of the phosphorene toward a weak acidic moiety as molecular iodine, similarly to what occurs for the $P_4$ moiety. Similar to $P_4$, the formation of P-I bonds should involve a structural rearrangement with a linear I---P-P disposition that has to be discharged in view of a limited degree of freedom, caused by the bent lone pairs orientation. In this regard, the only possibility is the initial adduct with the molecular iodine without any evolution.

Recently, we carried out a detailed analysis on the different coordination modes of unsaturated transition metal fragments, shown in Figure 4, or boranes with the phosphorene surface.[38] In particular, solid state calculations provided useful hints on the electronic, structural and energy features, ruling the phosphorene surface functionalization. The paper analyses in details the structural and the electronic effects (band gap) on the functionalized 2D material. The provided information will be the guidebook for future experimental tests in order to obtain new species with the desired properties.

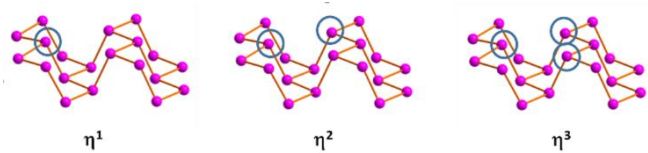

Figure 4: Possible coordination mode of unsaturated metal fragments on phosphorene surface.

## 4. High pressure behaviour of black Phosphorus layers

Black phosphorus has a crystalline structure (A17, C*mce*, Z=8). Phosphorene slabs are related to bP as graphene to graphite. Even if the synthesis, stabilization and functionalization of Phosphorene still represent very challenging tasks, the search for predicted and actually recently synthesized alternative single layer 2D Phosphorus allotropes has stimulated new theoretical and experimental investigations.[39] Within this perspective, pressure once again becomes relevant, because it represents a very effective tool to experimentally access rhombohedral A7 (R$\overline{3}$m, Z=2), another crystalline layered structure of P, which can be pictured as due to the stacking of blue phosphorene layers.[40]

Under room T compression A17 transforms indeed to A7 at ~ 5 GPa, which is stable up to ~11 GPa (Figure 5). According to literature, above this pressure A7 transforms to a non-layered metallic simple-cubic phase (sc, Pm$\overline{3}$m, Z=1), which remains stable up to 103 GPa. Above the latter pressure an incommensurately modulated structure (P-IV, IM, Cmmm(00g)s00) appears..[41-44] For pressure higher than 137 GPa and up to 282 GPa, P-IV transforms to a simple-hexagonal phase (P-V, sh, P6/mmm, Z=1). Above 262 GPa and stable up to 340 GPa a body-centered cubic structure (bcc), later identified as a superlattice structure (P-VI, cI16(I43d)), has been identified.[43]

Nevertheless, even if the room T phase diagram of P is known up to 340 GPa, a recent state-of-the-art high pressure synchrotron X-ray diffraction (XRD) experiment has shed new light about the low pressure range, with significant implications concerning the layered structures of P.[45] The XRD patterns, acquired during the room T compression of black phosphorus using a membrane Diamond Anvil Cell (DAC) with He as pressure transmitting medium, revealed indeed two unexpected peaks in the sc structure, appearing at 10.5 GPa and persisting up to 27.6 GPa, the highest investigated pressure.

The experimental observation demonstrates that a two-step mechanism rules the transition from the layered A7 to the non-layered sc structure, as suggested by theoretical predictions.[46] In the first step, rapidly occurring at 10.5 GPa, $\alpha$ approaches the sc limit and the rhombohedral unit cell becomes a slightly distorted sc one, but the atoms have not yet occupied the expected positions. In the second step, slowly occurring with pressure and not yet completed up to 27.6 GPa, the atoms slightly move toward the sc positions.

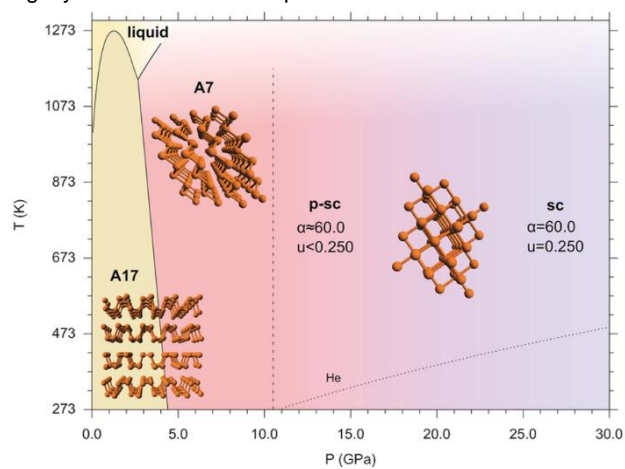

Figure 5: In the 10.0-30.0 GPa pressure range the phase diagram of elemental Phosphorus shows the regions of pressure and temperature in which A17 and A7 phases are stable. [42-44] In contrast with the reported data, where the phase

transition from A7 to sc is at 10.5GPa (marked by the dashed line), our data showed that at room the sc phase is not present up to 27.6 GPa and that between the A7 and the sc structures an intermediate pseudo simple-cubic (p-sc) structure exists (see text). The melting line of He is also displayed (dotted line). [44] In the figure the models of the structure are also displayed in the corresponding phases. (D. Scelta, A. Baldassarre, M. Serrano-Ruiz, K. Dziubek, A. B. Cairns, M. Peruzzini, R. Bini, M. Ceppatelli Interlayer Bond Formation in Black Phosphorus at High Pressure. *Angew. Chem. Int. Ed.* **2017**, *56*, 14135 –14140. Copyright Wiley-VCH Verlag GmbH & Co. KGaA. Reproduced with permission.)

Between from 10.5 GPa up to at least 27.6 GPa a *pseudo* simple cubic structure (p-sc) exists rather than the reported sc one, with three remarkable implications. First the higher pressure limit where the layered structures of P can be observed can be related with the presence of 3 shorter and 3 longer interatomic distances that structurally relates p-sc to layered A7. Second, the observation of the A7-like p-sc structure up to about 30 GPa brings order to the sequence of the high pressure limit of the A7 layered structure in group 15 elements, solving apparent anomalies and reconciling the high pressure structural behavior of P with those of heavier pnictogens.[47] Finally, as superconductivity in black Phosphorus is concerned, the existence of the p-sc structure up to ~30 GPa, provides new experimental evidence to account for the long debated anomalous pressure evolution of $T_c$ below 30 GPa.[48] The existence of the p-sc structure in P has been indeed explained by the delicate interplay between the strong *s-p* orbital mixing, dominant at low pressure, and the electrostatic contribution, which prevails under higher density conditions.[49] The identification of the key role played by the *s-p* orbital mixing in the stabilization of the layered structure opens new perspectives for the synthesis, stabilization and functionalization of new Phosphorene based materials and heterostructures.

## 5. Decorating bP with nanoparticles

The properties of exfoliated bP can be largely influenced by chemical and chemico-physical modifications of the surface. Thus, it is not surprising that by decorating the phosphorene surface with metal nanoparticles (NPs), a new material derives with different properties and therefore potentially useful for many applications in various fields, ranging from electronic and optical devices, to catalysis and finally to biomedical tools. Many papers on the functionalization of bP with NPs have appeared in the recent literature, here, a short summary of the most representative results is presented.

Pioneering work was achieved by the group of Scott Warren, who functionalized the surface of 2D bP with gold nanoparticles.[50] Starting from the complex [(PPh$_3$)AuCl] as precursor, in the presence of triethylamine (TEA) and light irradiation (λ = 465 nm), gold nanoparticles are formed on the surface of 2D bP with an average size of 20 nm. The mechanism of formation is attributed to the fact that being 2D bP a semiconductor, solar light induces the migration of electrons from the valence to the conduction band. These electrons are responsible for the reduction of Au(I) to Au(0) while the generated hole oxidizes TEA to TEA$^+$. The photocatalytic behaviour of this material are reported below.

Following this track, other groups have studied the functionalization of 2D bP with Au NPs and have discovered interesting applications. For instance, the work of Chu *et al.*[51] evidenced again the reducing capability of exfoliated bP where the synthesis of Au NPs takes place *in situ* in a very easy way, by mixing a suspension of HAuCl$_4$ and 2D bP in water.

Recently, Y.-N. Liu and his group[52] have functionalized bP nanosheets by *in situ* growth of Au NPs, using as precursor HAuCl$_4$ and the action of ultrasound. Interestingly, the authors discovered that the resulting nanohybrid Au@bP in the presence of ultrasounds, can produce singlet oxygen, $^1O_2$, which is highly cytotoxic and induces cell apoptosis, thus opening the way for its application in the sonodynamic therapy (SDT). Through *in vitro* and *in vivo* studies, it was shown that Au@bP can remarkably kill tumour cells and suppress tumour growth under ultrasound irradiation with few side effects.

Anchoring preformed Au NPs on the surface of bP nanosheets was performed by Y. Zeng[53] through polyetherimide chains (PEI) that act as a bridge between bP and NPs. Such hybridization simultaneously increases the $^1O_2$ production of bP nanosheets upon laser irradiation and also improves the light absorption of bP nanosheets promoting photothermal effects. As proof-of-concept, Au@bP-PEI resulted very effective in suppression of tumour growth both *in vitro* and *in vivo*.

The first report on the functionalization of 2D bP with platinum nanoparticles comes from the group of J. Kim.[54] The idea behind the incorporation of Pt NPs was the fabrication of a hydrogen sensor, taking advantage of the high capability of platinum to catalytically assist hydrogen dissociation. Mechanically exfoliated bP flakes were transferred onto source-drain electrodes (Ti/Au), preformed Pt NPs (3 nm size) were spin-coated on the surface of bP and subsequently, poly(methylmethacrylate) (PMMA) was deposited on top to prevent bP degradation. H$_2$-sensing performance was measured and a maximum sensitivity of 50% could be achieved, while pristine bP is insensitive to hydrogen.[55]

Soon thereafter, very similar results were obtained by Jung *et al.*,[56] who reported the surface functionalization of 2D bP with Pt NPs in solution. The gas sensing ability of Pt/bP was tested and resulted very high and selective towards H$_2$ gas, see Figure 6.

A completely different application for Pt NPs supported on bP was envisaged by Y.-N. Liu, who tested the nanohybrid Pt@bP in the photodynamic anticancer therapy (PDT).[57] The great potential of 2D semiconducting material in PDT is well known,[58] in particular the pioneer work of Y. Xie evidenced[59] the property of 2D bP, once irradiated with visible light, to generate singlet oxygen with high quantum yield. A drawback of the PDT therapy is the tumour hypoxia which hampers the *in vivo* efficacy by limiting the generation of reactive oxygen species (ROS) leading to poor therapeutic efficiency. This work demonstrates through *in vitro* and *in vivo* studies that Pt@bP can efficiently decompose the accumulated H$_2$O$_2$ in tumours to oxygen thus reducing hypoxia and thereafter improving the PDT activity.

Since in cancer radiotherapy high-Z elements as Au, Ag and Bi are of widespread use and they produce high-energy photoelectrons efficiently hydrolysing surrounding water molecules to produce ROS,[60] the group of X.-F. Yu achieved the growth of ultrasmall $Bi_2O_3$ nanoparticles directly on bP nanosheets aiming to further improve the efficiency of the 2D material in PDT. The new heterostructure showed an increased stability in water, being the NPs mostly located on the defect sites of bP.[61] *In vitro* and *in vivo* tests evidenced the synergistic effect of bP and $Bi_2O_3$ inducing an increased cell apoptosis. Thus, the bP-based radiosensitizers have huge clinical potential for synergistic cancer radiotherapy.

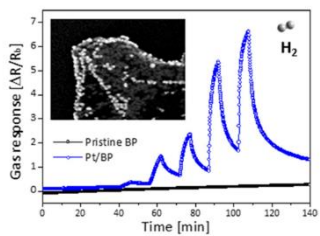

Figure 6. Comparison between pristine bP and Pt/bP towards $H_2$ gas sensing ability. Reprinted with permission from S.-Y. Cho, H.-J. Koh, H.-W. Yoo, H.-T. Jung *Chem. Mater.* **2017**, *29*, 7197.. Copyright 2017 American Chemical Society.

By reduction of $AgNO_3$ with excess sodium borohydride in the presence of bP, the nanohybrid Ag@bP was synthesized. TEM[62] revealed the average size of Ag NPs varies from 20 nm to 40 nm for 3wt% Ag@bP and 5wt% Ag@bP, respectively.

Spin polarised density functional theory (DFT) was used to characterized Ag and bP interaction. For low Ag content, a charge transfer from Ag to P and a localized electron distribution suggest a prevalent ionic nature of the Ag-P bond. On the other hand, in presence of a high amount of Ag, a covalent bond interactions are indicated by somewhat elongated Ag-P bonds due to the decreased charge transfer from Ag to P and more delocalized electron distribution. In conclusions, covalent bonds at the Ag/bP interface and Ag-Ag interactions are responsible of the strong stabilization of Ag NPs on bP.

The bP is a narrow-bandgap semiconductor, once it is functionalised with plasmonic Ag NPs on the surface, the resulting nanohybrid is supposed to adsorb a maximum amount of incident energy under resonance conditions. Thus, its photocatalytic activity was tested in a model reaction as the photodegradation of Rhodamine B in water. It was observed that increasing the size of Ag NPs from 20 to 40 nm and decreasing the thickness of the flakes, the reaction rate increased dramatically. The behaviour can be attributed to a change in the band gap which favours the formation of strongly oxidative holes and therefore enhanced the photoactivity.

Decoration of 2D bP with nickel nanoparticles was recently carried out by some of us following a different procedure.[63] Preformed Ni NPs[64] were immobilized onto the surface of freshly exfoliated bP. The new nanohybrid Ni/2D bP was fully characterized by surface techniques, as Raman, XPS, TEM, HAADF-STEM, confirming that 2D bP was chemically unaltered after functionalization and maintained its crystallinity. To evaluate the role of surface functionalization on the stability of the 2D material, a sample of Ni/2D bP was exposed to ambient humidity in the darkness and measured at regular intervals of time by TEM and Raman. A sample of pristine bP was kept in the same conditions and used as reference. While the latter decomposed to molecular compounds in two weeks, Ni/2D bP flakes appeared unchanged in one week time scale.[63]

Having 2D bP a higher surface to volume ratio than other 2D materials due to its puckered lattice configuration, we used Ni/2D bP as a heterogeneous catalyst in a model reaction, the semihydrogenation of phenylacetylene to styrene. High selectivity towards styrene was obtained at almost quantitative conversion, thus outperforming known catalytic systems based on supported Ni NPs.[65] The reuse of the catalyst was studied and after five successive runs the conversion and selectivity were unaltered, confirming the increased stability of the system in comparison to pristine bP.

A different approach was developed by the group of Pan and Zhang,[66] who synthesized the nanohybrid $Ni_2P$/2D bP in a one pot procedure, by thermal decomposition at 320°C of the organometallic precursor $Ni(acac)_2$ in presence of exfoliated bP and trioctylphosphine as "P" source. Nowadays, there is a high interest in the application of nickel phosphide as hydrogen generation (HER) electrocatalyst,[67] since Pt-based electrodes exhibit excellent catalytic performance but their large scale applications are limited by the high price and scarce metal reserve. Electrochemical tests were carried out dropping on the surface of the glassy carbon electrode an ink solution of $Ni_2P$/2D bP and interestingly the nanohybrid showed high HER performance, comparable to 20% Pt/C catalyst as shown in Figure 7, with low onset overpotential, small Tafel slope, high conductivity and good stability.

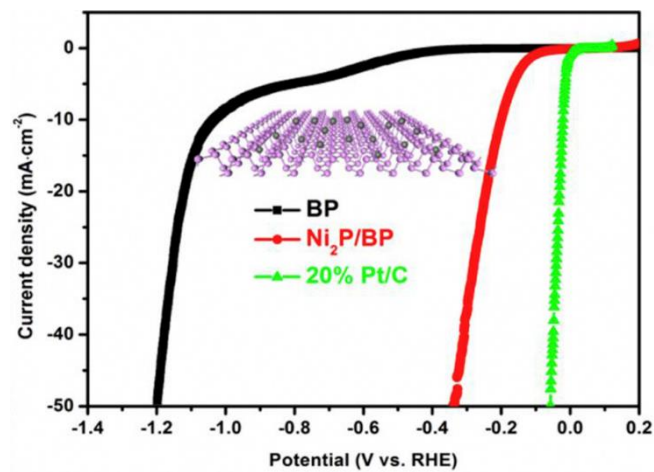

Figure 7. Linear scanning voltammetry curves for bP, $Ni_2P$/2D bP and 20% Pt/C catalysts. Reprinted with permission from Y. Lin, Y. Pan, J. Zhang *Int. J. Hydrogen Energy* **2017**, *42*, 7951 DOI: 10.1016/j.ijhydene.2016.12.030. Copyright 2017 Elsevier.

The photocatalytic activity of 2D bP, which has been already documented in processes as hydrogen production[68] and solar water splitting,[69] can be improved using bP nanosheets decorated with amorphous Co-P nanoparticles having 5 nm average size, were successfully isolated. Employed as photocatalyst, Co-P/bP exhibited robust hydrogen evolution from pure water achieving 5.4% energy conversion efficiency at 353 K and maintained the photocatalytic activity within 40 hours indicating high stability.[27]

## 6. Polymer-based nanocomposites containing 2D bP

In comparison to other 2D nanostructured systems, few studies have recently emerged on bP in contact with soft materials, such as thermoplastic polymers, for the preparation of nanocomposites. These polymer-based hybrids are generally prepared with the dual objective to stabilize the bP-nanoflakes (or phosphorene), which is mandatory for any application, and to design devices for optoelectronics with improved performances.

In fact, the greatest challenge regarding bP, especially once exfoliated, comes from its fast degradation when exposed to ambient conditions. Indeed, the reaction with oxygen in the presence of light governs the degradation of 2D bP (up to 20-30 nm thickness). Reactive oxygen species (ROS) are generated depending on different environmental factors and are responsible of 2D bP photo-oxidation extent as discussed in the literature, where degradation mechanism and the use of different chemicals as effective ROS quenchers[70-74] have been proposed. Passivation of bP nanoflakes has been achieved through suitable coatings such as layered materials (hexagonal Boron Nitride, h-BN),[71] oxides ($Al_2O_3$),[72] or polymers. Among these, polymer coating or embedding appear to be the easiest, with particular reference to the use of poly(methylmethacrylate) (PMMA),[73] polystyrene (PS)[74] and polyaniline (PANI),[75] which efficiently preserve the previously arranged bP nanoflakes (vide supra).

Most of nanocomposites have been prepared by mixing the polymer with already exfoliated bP generally obtained by sonication-assisted liquid-phase exfoliation (LPE) methodologies involving the use of different solvents/exfoliating agents. DMSO,[71] NMP,[76,77] N-cyclohexyl 2-pyrrolidone (CHP),[78] saturated NaOH/NMP solution,[79] aqueous solution containing surfactants[80] or other less eco-friendly chemicals as N,N,-dimethylformammide (DMF),[81] have been used for such purpose. The obtainment of bP-nanoflakes is, in these cases, generally a heavy, multi-step, procedure comprising sonication, centrifugation, isolation of the target material and, above all, removal of non-volatile and non-friendly solvents, which is often a drawback for the fabrication and the performances of devices.

In spite of this problematic procedure, appropriate suspensions of phosphorene can be easily mixed with polymer solutions to obtain hybrids for a variety of applications. For example polycarbonate (PC),[78] polyvinyl alcohol (PVA)[76,77] and polyimide (PI)[77] (all transparent and flexible polymers) have been used to prepare saturable absorbers (SA) for Q-switched fibre lasers whose performances are completely comparable with those obtained by employing mechanically exfoliated bP. In addition, the polymer works not only as a carrier but also as a protector of bP nanoflakes (acting as SA) from environmental deterioration, by passivating their surfaces. The Q-switched laser fibres have been designed and realized [76,77] in a similar arrangement. In both cases it has been demonstrated that the enveloping of 2D bP in polymer matrices provides an alternative solution to manage the environmental degradation of nanoflakes and to enhance/stabilize the long-term performances of bP as suitable Q-switcher candidate.

Polymers can be also used as "stabilizers" during and after the exfoliation of bP; an ethanol solution of polyvinylpyrrolidone (PVP)[82] has been successfully employed to yield stable dispersions constituted by a large fraction of single bP-nanosheets. Such PVP-stabilized phosphorene with enhanced stability towards oxidation, compared to nanosheets prepared by LPE in conventional solvents, exhibits once again, saturable absorption characteristics, suggesting potential applications as ultrafast broadband absorbers. In addition, this material can be easily subjected to decoration with metal nanoparticles (in particular Cu nanoparticles) whose electrocatalytic activity, appears superior to that of the graphene-analogous material, increasing the interest in studying the possible application of such transition metal/phosphorene heterostructures in the energy conversion field.

By considering the peculiar electronic properties of phosphorene, its embedding in semiconductor polymers, such as polypyrrole (PPy),,[79] polyanilyne (PANI)[81] and (poly(3,4-ethylene dioxythiophene):poly (4-styrenesulfonate) (PEDOT:PSS)[83] can be considered as a possible strategy for future electronic applications.

Semiconductor composites and capacitors with good processability and filmability have been prepared starting from preformed soft materials or their monomers. In a very schematic description of methodologies used to prepare these devices, the first step involves the bP delamination by LPE followed by mixing with the polymers or, in a few cases, with the monomers. Zhang and co-worker[82] prepared mixed dispersions of bP-PEDOT:PSS with different weight ratios by sonication-assisted LPE in isopropanol containing the PEDOT:PSS polymer.

In addition, PPy/bP composites have been prepared by electrochemical deposition[79] starting from pyrrole monomer contained in the electrodeposition solution with sodium dodecylbenezenesulfonate and p-toluenesulfonic acid. The methodology here adopted is suitable to fabricate self-standing PPy/bP laminated films directly attached onto indium-tin oxide (ITO) and self-assembled into a flexible conductive film. Luo et al.[79] claimed that the existent bP nanosheets lead to a laminated assembly avoiding dense and disordered stacking of PPy films. This result suggests that, by synthetizing the composite during the polymerization of the monomer, the phosphorene is able to address/orient the polymer chains growth promoting the self-assembly of the hybrid material: This architecture is especially performant for applications (figure 8).

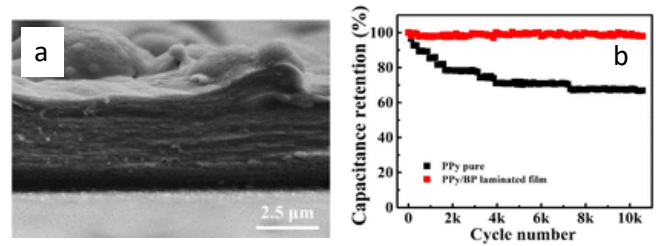

Figure 8. a: Cross-section SEM images of PPy/bP laminated film, b: capacitance retention of laminated and pure films after 10000 cycles under a current density of 5 A g$^{-1}$. Reprinted and adapted with permission from Figure 2f and 3e of S. Luo, J. Zhao, J. Zou, Z. He, C. Xu, F. Liu, Y. Huang, L. Dong, L.; Wang, H. Zhang *ACS Appl Mater Interfaces* **2018**, *10*, 3538. Copyright 2017 American Chemical Society.

In fact, this laminated structure leads to a device with high capacitance (497 Fg$^{-1}$) while cycling stability shows no deterioration even after 10000 charging/discharging cycles (figure 8b). From these results, it seems that the 2D bP acts as a template and induces, during the electrodeposition, an ordered configuration of PPy. This polymer/nanostructure architecture is probably able to hinder the dense PPy stacking and creates wrinkles and/or holes, which should promote the electrolyte diffusion thus improving the energy storage of film electrode.

Such capability of phosphorene to direct the morphology of a polymer (similarly to that of other 2D material like graphene) has been even proved by deposing block copolymers onto a substrate of 2D bP.[84] The asymmetric polystyrene-block-polymethyl methacrylate block copolymer (PS-block-PMMA) thin film results self-assembled to form perpendicular orientation of sub-10 nm PS nanopore arrays normal to the interface. This procedure has been claimed as a useful methodology to obtain nanostructural microdomains serving as promising template for surface patterning of bP-nanoflakes. Notably, a similar approach is currently under investigation in our laboratory as a tool to design an electrical percolation path between 2D bP nanosheets.

Even if it is not clearly explained, these effects can be ascribed to the generation of interfaces where the structural characteristics of polymers are led (or at least modified) and the properties of bP-nanoflakes strengthened, especially thanks to the improved environmental stability. Closer interactions between the polymer and the 2D bP can be granted by growing the polymer chains near or from the phosphorene layers as, for some aspects, in the case of the bP-PPy nanocomposites above discussed.[79] A fabrication approach, for the first time mentioning the *in situ* polymerization, as suitable methodology to establish synergistic effects between bP-nanoflakes and polymer matrix has been reported by Pumera *et al*[81] who have designed high-performance pseudo capacitors based onto PANI.

The experimental methodology is laborious and requires the use of solvents, inert atmospheres and different reagents/apparatus (see details in ref 81) owing to solution polymerization of aniline in the presence of suspension of already exfoliated bP. However, the preparation of bP/PANI composites with a co-continuous phase between the PANI and bP-nanoflakes have been evidenced.

2D bP with large surface area serves as a substrate for the polymerization and grants the uniform growth of PANI allowing for efficient charge storage with enhanced ion transport. Indeed, bP/PANI composites exhibit outstanding capacitive performance of 354 Fg$^{-1}$ at a current density of 0.3 Ag$^{-1}$, excellent rate capability and improved long-term cycling stability with reference to PANI (produced with the same methodology without phosphorene) and starting exfoliated bP, definitely confirming the synergistic effects established owing to interfacial interactions.

Once the *in situ* polymerization had been established as a promising tool to synthetize nanocomposites containing 2D bP with enhanced interfacial interactions, new advances have been provided by Passaglia *et al.* [85] by combining LPE of bP in the monomer with *in situ* polymerization, in the bulk, without using additional solvents (figure 9).

MMA, styrene (Sty) and N-vinyl 2-pyrrolidone (NVP) have been successfully used as solvent/exfoliating agents of bP by ultrasonication in ambient atmosphere and after insufflation of nitrogen and the addition of a radical initiator (AIBN) the polymerization have started.

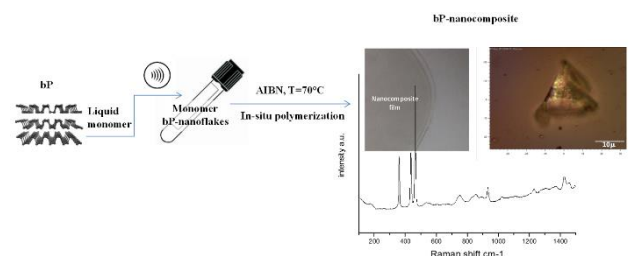

Figure 9: Schematic representation of bP nanocomposites preparation by sonication assisted bP-LPE in the monomer (NVP) followed by *in situ* polymerization; picture of film, optical microscopy of a portion of film showing a bP-nanoflakes aggregate and Raman spectrum of composite showing the diagnostic signals of bP (the A$_g^1$, B$_{2g}$, A$_g^2$ modes) together with bands associated to the polymer matrix (see ref 85). Reprinted with permission and adapted from Figure 2f and 3e of E. Passaglia, F. Cicogna, F. Costantino, S. Coiai, S. Legnaioli, G. Lorenzetti, S. Borsacchi, M. Geppi, F. Telesio, S. Heun, A. Ienco, M. Serrano Ruiz, M. Peruzzini *Chem. Mat.* **2018**, *30*, 2036. Copyright 2018 American Chemical Society.

PMMA-, PS-, PVP- based nanocomposites containing 2D bP have been collected in good yields. This very simple procedure allows the design of composites with different polymer/bP ratios, and the phosphorene has been demonstrated to be long-term stabilized against oxidation even after processing the composites by compression moulding (at 180 °C), solubilisation in different solvents (anisole, chloroform, water) and aging by UV-irradiation. AFM characterization confirms the presence of aggregates where the 2D bP flakes result enveloped in polymer chains presumably growing from the phosphorene surfaces activated by AIBN, which has been recently proved to be able to react with phosphorene surface.

This really stable material has been successfully used as a platform suitable to engineer the properties of phosphorene and to exploit its potential in the fabrication of devices.[85] A bP-PMMA composite containing about 0.2 wt% of exfoliated bP solubilized in anisole has been spin-coated in ambient condition in a cleanroom environment on SiO$_2$/Si pre-patterned substrates. By analysing the sample (figure 10a and b) single individual nanosheets have been observed, confirming their structure being preserved for at least 3 months.

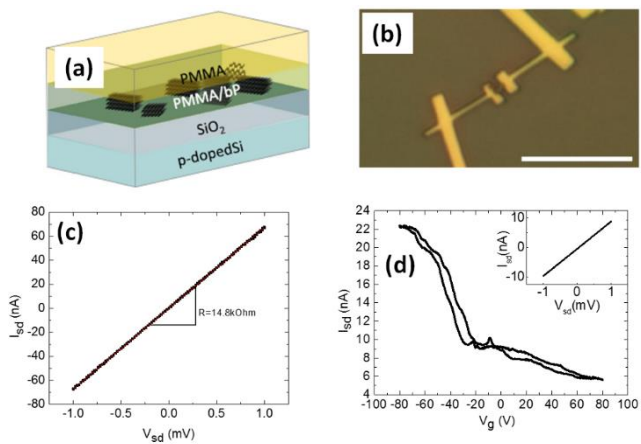

Figure 10. (a): bP nanosheets embedded in PMMA, on a Si/SiO$_2$ substrate.; (b) Optical microscopy after device fabrication. The scale bar is 10 µm; (c) Current (I) versus voltage (V) characteristics of the device at room temperature. Resistance R = 14.8 kOhm. (d) Low temperature measurements at 4.2 K. Source-drain current as a function of gate voltage (V$_{sd}$ = 1 mV) shows p-type behaviour. Field-effect mobility 35 cm$^2$ V$^{-1}$ s$^{-1}$. I$_{sd}$ vs. V$_{sd}$ curve, shown in the inset, gives 108 kOhm. Adapted under CC BY 3.0 licence from Figure 2f and 3e of F. Telesio, E. Passaglia, F. Cicogna, F. Costantino, M. Serrano-Ruiz, M. Peruzzini, S. Heun *Nanotechnology* **2018**, *29*, 295601 DOI: 10.1088/1361-6528/.

The selected nanosheets have been then processed by standard fabrication techniques (see ref [85] for details) by covering with PMMA and using the electron beam lithography (EBL). This procedure is able to grant that the bP flakes are confined close to SiO$_2$ surface. After the lithographic process, the sample has been processed without the need to protect the flake and the device as for example by using a glove box or any other kind of protective atmosphere. Finally, the electrical transport properties of the device have been measured in vacuum both at room and at low temperature (figure 10c and d) showing a resistivity and carrier mobility (35 cm$^2$ V$^{-1}$s$^{-1}$) which are characteristic of black phosphorus and stable over at least 50 days. In addition the expected p-type behaviour upon gate voltage modulation has been confirmed.

## 7. General transport properties of bP

Keyes[86] and Warschauer[87] showed that bP is a p-type semiconductor in the 50's and the 60's. In 1983, the group of Narita succeeded in an n-type doping of bP by Te.[88] In the same paper, also a high hole mobility of bP of up to 6.5 x 10$^4$ cm$^2$/(Vs) was reported, with a strong anisotropy between its three crystallographic directions. Also a strong anisotropy of the effective masses of carriers in bP was reported.[89] Finally, in 2014 the interest in bP exploded [90] when it was demonstrated that bP can be exfoliated down to a few layers.[9, 14, 91-93] The interest was fuelled by the fact that bP offers several unique properties which will be discussed in the following. An exciting and potentially very useful feature of bP is its high-mobility transport anisotropy. A hole mobility of 10,000 cm$^2$/(Vs) was predicted for monolayer bP,[94] together with a strong in-plane anisotropy due to the puckered structure of the bP layer. First devices showed a peak field-effect mobility of ~1000 cm$^2$/(Vs) at room temperature for 10 nm-thick bP flakes,[9] already twice as high as typical values for commercial Si devices.[95] Mobility is approximately constant below 100 K and decreases due to electron-phonon scattering (µ∝T$^{-0.5}$) above 100 K.[9,92] Recent experiments reported mobility values up to 45,000 cm$^2$/(Vs) at cryogenic temperatures for few-layer bP encapsulated in hexagonal BN (h-BN).[96]

The structural anisotropy of the bP layer reflects in anisotropic physical properties. An anisotropic behaviour between the armchair (ac) and the zigzag (zz) directions of bP has been observed in infrared spectroscopy,[92] Raman spectroscopy,[9, 97] DC conductivity $\sigma$,[92] thermal conductivity $\kappa$[98,99] and Hall mobility.[92] Once calibrated, these methods can be employed to identify the armchair and zigzag directions of a given bP flake. We note that the anisotropy is not always in the same direction. For example, the electrical conductivity along the armchair direction is larger than along the zigzag direction, with $\sigma_{ac}/\sigma_{zz}$~1.5,[92] while for the thermal conductivity, the opposite is true: it is larger along the zigzag direction, with $\kappa_{ac}/\kappa_{zz}$~0.5 at 300 K,[98, 99] orthogonal to the preferred direction of electrical conduction.

By scanning tunnelling microscopy and spectroscopy (STM and STS) it has recently been shown that bP is p-type due to P single atomic vacancies which induce acceptor states in the bP band gap near the valence band edge.[100] Controlled doping of few-layer bP flakes has been achieved by surface transfer doping, where atoms or molecules at the surface of the flake accept or donate electrons. In an early paper, Cs$_2$CO$_3$ has been found to be a strong electron-dopant, while MoO$_3$ is a hole-dopant.[101] Also other molecules have been proposed for electron and hole doping via molecular charge transfer.[102] Te-doped bP has been synthesized and exfoliated,[103] the resulted flakes showed a p-type behaviour, in contrast to the previously mentioned n-type doping demonstrated in the 80s.[88] The authors attributed this to the low amount of Te used in this study. Similarly, Se-doped bP showed p-type behavior, as well.[104] It was predicted that group I metals (Li, Na, K) transfer a large amount of charge to the bP and thus are n-dopants.[32, 105] This was experimentally confirmed for Lithium[106] and Potassium.[107] Also evaporation of Cu on the surface of a few-layer bP flake leads to an n-type doping of bP with the Fermi level close to the conduction band minimum.[91]

The surface transfer doping of bP by potassium has been shown to induce a strong vertical electric field, which, through surface Stark effect, modifies the band structure of bP from semiconductor to band-inverted semimetal. At the critical point of this band inversion, two Dirac points are created, with linear dispersion in armchair and quadratic dispersion in zigzag direction.[107] It has been predicted that this will also have a deep impact on the low-temperature magneto-transport properties of bP. In the absence of a vertical (perpendicular) electric field, Landau levels are equidistant (as expected for a semiconductor with parabolic dispersion), while the emergence of a pair of Dirac points makes them follow a $\sqrt{B}$ characteristics (as observed in graphene).[108]

Electron-doped monolayer bP has been predicted to be a superconductor provided that the doping level is high enough,[109] with a critical temperature as high as 17 K.[106] In fact, superconducting bP has been recently demonstrated by alkali metal intercalation.[110]

## 8. 2D bP stability, optimization, and device properties

### 8.1. Environmental Stability of bP

For the reasons outlined in Section 2, the method of choice for bP device fabrication is still mechanical exfoliation. An additional difficulty compared to graphene is that bP is not stable in air but degrades quickly within days or weeks.[14] This is due to a photo-oxidation reaction in which water and oxygen from the environment react with the bP flake in presence of light to form $PO_x$ species.[111] Therefore, even if the device fabrication is entirely performed under a protective environment (for example, inside a glove box), new and effective bP passivation strategies are to be developed, in order to be able to expose the final device to ambient conditions.

Several groups have shown that an overlayer of $AlO_x$, grown by atomic layer deposition (ALD), can effectively suppress device degradation.[112,113] Another, easier, approach is to encapsulate the final device in PMMA,[114] which has actually been used also in a wider context. As discussed in Section 6, bP can be exfoliated in MMA and remains then protected in a PMMA matrix, which has been spin coated on a Si/SiO₂ substrate and used for device fabrication.[85] Covalent functionalization of bP has been proposed to passivate the surface of bP flakes and shown to protect bP from degradation, with improved field-effect transistor (FET) characteristics.[32] Similarly, decoration of exfoliated bP with Ni nanoparticles has been shown to stabilize bP flakes for at least one week[63]. For further details, see Section 5.

Perhaps the best way today to protect bP flakes from oxidation is to encapsulate them in h-BN,[115] This not only maintains the bP stable, but allows also for highest mobility: the devices with the highest carrier mobility are based on h-BN/bP/h-BN heterostructures[96]. For the assembly of such sandwiched bP heterostructure devices, usually the van der Waals transfer technique is employed, in which individual layers are one-by-one picked up by a PDMS stamp before the full stack is finally deposited onto a substrate.[116]

### 8.2. Controlled Thinning of bP flakes

The thermal stability of bP was investigated by several groups.[117] bP sublimation starts at around 375 °C and proceeds in a layer-by-layer fashion, starting from vacancies in the surface. These vacancies grow with an elliptical shape, with their long axis aligned along the zigzag direction of the bP crystal, see Figure 11. Based on these results, a method for the thermal thinning of few-layer bP flakes has been proposed and demonstrated.[118]

Another approach was suggested by Pei *et al.* based on surface oxidation,[119] through the exposition of a bP flake to an oxygen plasma. With exposure time, the thickness of the oxide increases, but at the same time, the plasma sputters away the topmost layer of the oxide, so that finally a dynamical equilibrium is created, resulting in a constant oxide thickness during the process. To further improve the stability of the film, the samples were finally capped with a protective $Al_2O_3$ layer grown by ALD. This treatment has improved the optical [119] and electric [120] properties of bP.

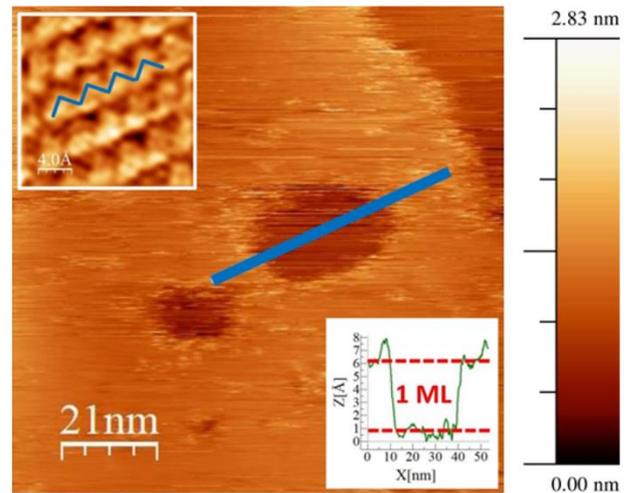

Figure 11: STM image of craters on a bP flake, annealed at 400 °C for 5 mins, measured at room temperature. The upper left image shows an atomic resolution zoom-in. Craters are aligned along the zigzag direction indicated by the blue lines. The lower right inset shows that the crater is monolayer-deep. Scan parameters: 1.5 V / 100 pA.

### 8.3. Field-effect Transistors

The first bP FET was demonstrated by Li *et al.* [9] with a drain current modulation (on-off ratio) of $10^5$. In back gate sweeps, a minimum of conductance was detected around $V_g = 0$ V, with increasing conductance on either side. Hall measurements showed that this was related to an inversion in carrier sign, with negative (positive) gate voltages corresponding to hole (electron) conduction. I-V curves were linear (Ohmic) in the hole branch, while they were strongly nonlinear on the electron side, which was explained with the formation of a Schottky barrier between the semiconductor (bP) and the metal contacts (Cr/Au). Similar FETs have been shown to be useful also in high frequency applications, and a cutoff frequency of 12 GHz has been demonstrated. [121] A more sophisticated device with dual (top and bottom) gates was shown to work as a velocity-modulated transistor, with the possibility to independently tune one top and one bottom FET channel with equal or opposite polarity, *i.e.* a vertical p-n junction could be created.[122] In a similar device, the ambipolar bP double quantum well was embedded between two layers of h-BN, which resulted in improved carrier mobility and the observation of the quantum Hall effect.[123] Local electrostatic gating was used to create a lateral p-n junction in bP and to observe the photovoltaic effect, useful for energy harvesting in the near-infrared.[124] Towards more complex integrated circuits based on 2D materials, CMOS logic circuits containing few-layer bP were demonstrated. CMOS uses complementary and symmetrical pairs of p-type and n-type FETs for logic functions. In 2014, a CMOS inverter with bP for the p-type transistor was build. Since at that time no n-type bP was available, for the complementary n-type transistor MoS$_2$ was employed.[10] Two years later, selective deposition of Cu on part of a bP flake was used to create n-type channels in an otherwise intrinsically p-type material, yielding a 2D complementary inverter device entirely based on bP.[125] FETs on bP were also realized on flexible polyimide substrates, achieving a carrier mobility in excess of 300 cm$^2$/(Vs), more than five times higher than previous flexible transistors. An AM demodulator (useful for radio receivers) was demonstrated.[126] This opens the door for the application of bP in flexible electronics. Recently, bP has also proposed as the ideal material for tunneling-FETs, due to its direct band gap, low effective mass, anisotropic bands, and high mobility.[127] However, an experimental demonstration of this type of device is to our knowledge still to be done.

**8.4. Low-temperature magneto-transport**

These studies prepared the ground for low-temperature magneto-transport experiments on few-layer bP. In 2014, four groups independently observed quantum (Shubnikov-de Haas) oscillations in bP thin films which demonstrated two-dimensional transport in these samples.[17c,114,116,128] Some of these studies were already based on devices encapsulated in h-BN. This technology has later allowed the observation of the quantum Hall effect in bP, a breakthrough experiment which was only possible thanks to a hole mobility as high as 6000 cm$^2$/(Vs).[129] Very recently, even the observation of fractional quantum Hall features in bP was reported.[130]

Weak localization was studied in few-layer bP FETs,[131-133] see Figure 12(a). Weak localization is a quantum-mechanical effect related to phase-coherent scattering at low temperature[134] and results in a resistance increase due to enhanced backscattering, as shown in Figure 12(b). Application of a magnetic field destroys time-reversal symmetry and cancels the contribution of weak localization to the magneto-resistance. The resulting data can be analysed by the theory of Hikami *et al.*[135] Dephasing lengths $L_\varphi$ between 30 nm and 100 nm were found, which exhibit a strong dependence on carrier density and temperature (Figure 12(c)). The temperature-dependence is controversially discussed, but there is evidence for a deviation from the expected $L_\varphi \propto T^{-1/2}$ behaviour characteristic of electron-electron scattering in the presence of elastic scattering in two dimensions,[131,132] as illustrated in Figure 12(c).

This was explained by a quasi-one-dimensional behaviour of bP due its pronounced crystalline anisotropy.[132] The relevance of crystalline anisotropy was also highlighted in a recent in-plane magneto-transport study which showed an unexpected, non-classical longitudinal magneto-resistance.[136] A transition from weak to strong (Anderson) localization was observed in bP upon reduction of the carrier concentration.[137]

**8.5. Thermoelectrics**

Thermoelectric applications are based on the Seebeck effect which describes the conversion of heat flow (due to temperature gradients) into electrical energy. Whenever there is a temperature gradient in a conductive material, a potential difference $\Delta U$ measured under open-circuit conditions, $\Delta U = -S \cdot \Delta T$, with $S$ the Seebeck coefficient of the material. The performance of a material as a thermoelectric generator is described by the dimensionless thermoelectric figure of merit $ZT$, which is defined as $ZT=\sigma S^2 T/\kappa$, where σ is the electric and κ the thermal conductivity. Therefore, in order to optimize $ZT$, a material should display high electrical and low thermal conductivities. Here, the pronounced crystalline anisotropy of bP is advantageous, because along the armchair direction, the ratio $\sigma/\kappa$ is three times larger with respect to the zigzag direction. In fact, a $ZT$ of 2.5 at 500 K has been predicted for bP, [138] while the experimentally observed values are still much lower.[99] This might be improved by going to bP nanoribbons.[139]

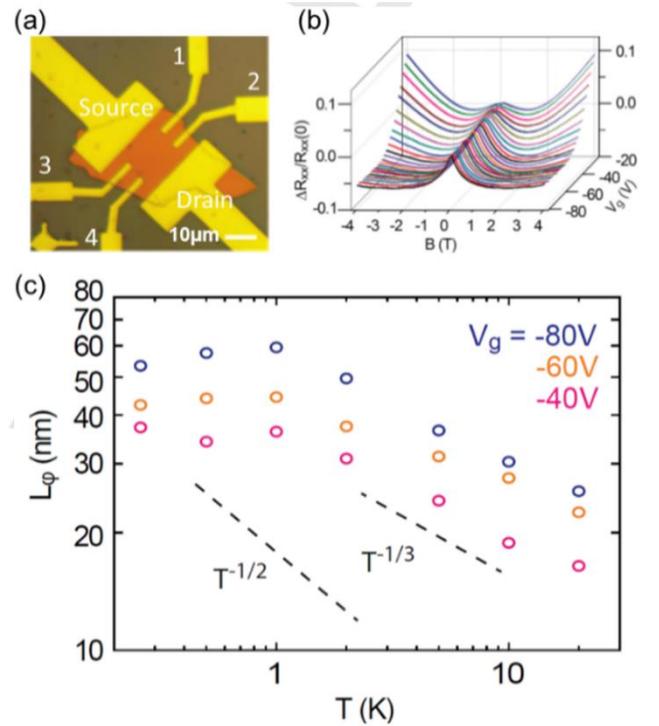

Figure 12: (a) Optical reflection microscopy image of a bP FET. The longitudinal resistance R$_{xx}$ is measured with voltage probes 1 and 2. (b) Weak localization measurements. A weak localization peak is observed in this plot of the normalized longitudinal resistance (R$_{xx}$(B) − R$_{xx}$(0))/R$_{xx}$(0) versus magnetic field B and gate voltage Vg at T = 0.26 K. (c) The inelastic scattering length L$_\varphi$ versus temperature T at various gate voltages Vg. The T$^{-1/2}$ temperature dependence associated with electron-electron scattering in the 2D diffusive limit is shown for comparison, as well as a T$^{-1/3}$ power law. Figure adapted by N. Hemsworth, V. Tayari, F. Telesio, S. Xiang, S. Roddaro, M. Caporali, A. Ienco, M. Serrano-Ruiz, M. Peruzzini, G. Gervais, T. Szkopek, S. Heun *Phys. Rev. B* **2016**, *94*, 245404 DOI: 10.1103/PhysRevB.94.245404 under the CC 4.0 licensee.

**8.6. Gas Sensors**

Due to their large surface-to-bulk ratio, two-dimensional (2D) materials are intensely investigated for gas sensing applications. With respect to competing 2D materials like graphene and MoS$_2$, bP has the advantage of a direct band gap (which can be tuned by layer number) and a high carrier mobility. Therefore, bP is a very promising material for sensing applications. Using density functional theory (DFT), the adsorption of various molecules on bP was investigated, and it was found that NO and $NO_2$ have the strongest binding to bP, indicating that bP would be a good sensor for these toxic gases.[140] A charge transfer mediates the binding between molecule and bP, i.e. molecular doping of the bP, which results in a change in resistance of the film, which can be measured.[140] These theoretical predictions were verified using bP FET devices [141,55] that reached sensitivity for $NO_2$ detection of 5 ppb.[141] The sensitivity for other gases, including $H_2$, was much lower, underlining the high selectivity of the sensor to $NO_2$.[55] Obviously surface transfer doping is more efficient for thin bP flakes, but for the thinnest flakes, the increase in band gap and the related reduction in carrier concentration leads to a lower binding energy for the molecules, which results in an optimum bP flake thickness of ~ 5 nm for $NO_2$ sensors.[55] In an experimental comparison between $NO_2$ sensors based on different 2D materials, bP was clearly outperforming $MoS_2$ and graphene.[142] bP can be functionalized to be sensitive to other gases, different from NO and $NO_2$. A relevant example for applications is $H_2$, for which a bare bP sensor is not performing, while Pt-functionalized bP provides much better results.[143,56] Long-term stability of the sensor of more than 4 weeks was reported, which was attributed to the functionalization of bP with the noble metal.[56]

bP has also been successfully used for humidity sensing,[144, 145] as a sensor for vapor (methanol),[146] for ion sensing,[147] and as a detector for the cardiovascular disease biomarker myoglobin.[80] A good overview of the relevant literature is included in the review paper by Gusmao et al..[148]

## 9. 2D bP applications in optoelectronics, photonics and photocatalysis

### 9.1. General opto-electronic properties

The thickness-dependent direct electronic bandgap of bP makes it a promising material also for nanophotonic and optoelectronic applications, [92,124,149-153] filling the gap left by other well-known 2D materials such as graphene (zero band-gap) [12,154] and group IV transition metal dicalchogenides (which show responsivity in the UV-vis spectral range).[155] In addition, the electronic band structure of phosphorene can be changed through the local mechanical modification of its geometry [156-159] allowing an even broader spectral range coverage. Thanks to its crystalline structure, phosphorene shows also optical anisotropy that is correlated with the in-plane electrical anisotropy. Regarding light absorption, optical selection rules render the extinction coefficient along armchair higher by 20% with respect to that in the zigzag direction.[160,161]

Phosphorene also shows anisotropic properties in its complex refractive indices in the spectral range between 480 and 650 nm, as easily obtained from anisotropic optical contrast spectra elaborated with the Fresnel equation,[162] giving to phosphorene the light modulation properties of linear polarizers, phase plates and optical compensators. This allows to easily integrate phosphorene into multilayer structures such as photonic crystals and microcavities[163,164] which can be applied as optoelectronic components with switching, filtering and tailored emissive properties.[165-168]

Regarding emission properties, those of phosphorene are largely dominated by excitons. This is due to the large exciton binding energy ($E_b$) which gives high stability to excitons in phosphorene. Excitonic emissions of phosphorene with varying thickness and band gap have been studied through photoluminescence(PL) spectroscopy, as shown in Figure 13.

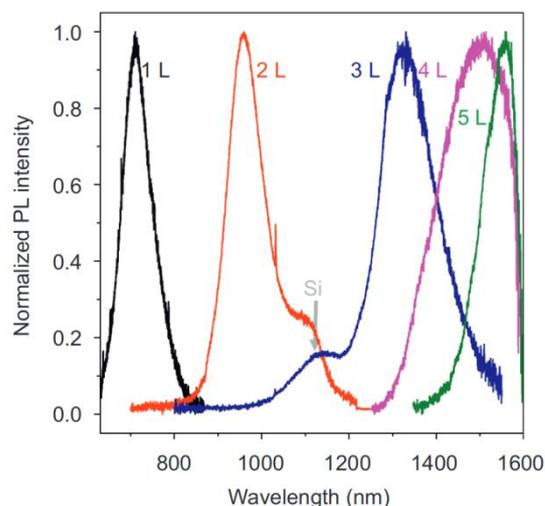

Figure 13. Normalized emission spectra of 2D bP samples with different thickness. Reprinted with permission and adapted from J. Yang, R. Xu, J. Pei, Y. Myint, F. Wang, Z. Wang, S. Zhang, Z. Yu, Y. Lu *Light: Sci. Appl.* **2015**, *4*, e312 DOI: 10.1038/lsa.2015.85. Copyright 2015, Springer Nature.

The blue-shifting of the PL peak n with the decrease of the 2D bP thickness is a clear indication that the bandgap ($E_g$) increases. The peaks broadness is due to the synergic effects between quantum the layers interaction and quantum confinement. [169,170] Together with the blue shifting, the peak intensity also increases for the thinner 2D bP samples, due to their lower density of states, at the valley edges,[9] which increases in turn the relaxation rates of separated charges and enhances the efficiency of their recombination process. The electronic bandgaps are higher than the optical gaps, so obtained, due to the above mentioned high $E_b$, calculated to be as high as 0.76 eV [169,171] in isolated phosphorene monolayer, and can be strongly reduced by screening from adjacent layers in devices (*i.e.* the dielectric layer, protecting agents and electrodes in FETs), or enhanced by uniaxial transverse strain.

The high stability of excitons also allows the existence of other excitonic states, likecharged excitons (*i.e.* trions, arising from the coupling between a free charge and an exciton), whose investigation in TMDs 2D materials has already given fundamental contributions to the study of spin or charges multiplication, many-body coupling, *etc*.[172-174] It has been also demonstrated that it is possible to modulate the trions in a FET electrostatically or through photoinduced charges injection, as observed by the change in the relative intensity of the exciton (~1100 nm) and the trion (~1300 nm) emissions.[175-176] Like excitons, also trions have been demonstrated to have a high $E_b$, much larger than other 2D systems and comparable to 1D materials like carbon nanotubes.[177] This is due to the confinement of trions and excitons in a 1D-like dimension, leading both the exciton and trion emissions to be anisotropically polarized along the same crystalline direction of phosphorene.[175]

By introducing *sub*-band gap electronic states through defects. the photo-response of 2D bP can be tuned. For example, bridge-type intrinsic surface oxygen defects may act as recombination centers to generate photons emission.[178] Indeed, when phosphorene is over-exposed to $O_2$/ozone plasma treatments, two new intense PL features arise at 780 nm and 915 nm, attributed to the excitons localized at two oxygen defects with different geometries respectively.[119] Controlled oxidation of 2D bP also allows to obtain different phosphorene oxides ($P_4O_n$), with 2D planar and 1D tubular forms, with crystalline or amorphous domains depending on the fabrication process, and with bandgaps monotonically increasing from the visible to the deep UV depending on the O-to-P ratio. For example, focused laser beam allowed to obtain, through photo-oxidation, suboxides with well-defined micropatterns on the 2D bP surface, with bandgaps easily tuned from ~2.3 eV to ~5.1 eV through the control on the laser power[175,176] and enhanced photoresponsivity in the corresponding device.[152]

## 9.2. Optoelectronic and photonic components and devices

Due to its exciting properties, 2D bP has been employed as an efficient active material in many optoelectronic devices. Thanks to the tuneable band gap and to the resonant-inter band transition between highly nested valence and conduction bands, combined to its high charge mobility, 2D bP has been proven as an excellent active material for light sensing and optical imaging. In the UV region, 2D bP-based phototransistors were demonstrated to reach a specific detectivity of $\sim 3 \cdot 10^{13}$ Jones and a photo responsivity as high as $\sim 10^5$ A W$^{-1}$ at an applied source-drain voltage of 3 V, representing a record responsivity within 2D materials-based FETs.[151] On the other hand, in the vis-to-MIR (visible to mid infrared) spectral region, 2D bP-based phototransistors have shown a lower photo-responsivity (up to ~80 A W$^{-1}$),[179] which is nonetheless accompanied by a good dynamic bandwidth (kilohertz range), a high speed (~ms) and broadband response,[150] demonstrated in 2D bP-based single-pixelphotodetectors[180] and photodetector arrays[181] with high (diffraction limited) optical imaging resolution(Figure 14).

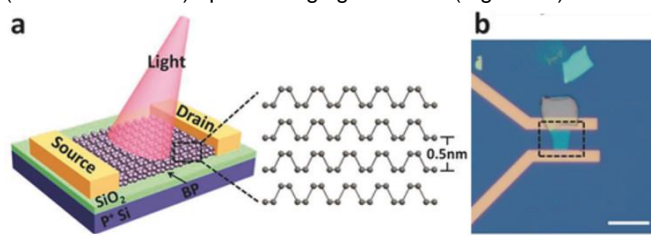

Figure 14. Schematic (a) and microscope (b) images of the 2D-bP-based photodetector. Reprinted with permission and adapted from J. Miao, B. Song, Z. Xu, L. Cai, S. Zhang, L. Dong, C. Wang *Small* **2018**, *14*, 1702082 DOI: 10.1002/smll.201702082. Copyright 2017 Wiley.

The integration of 2D bP with a silicon photonic waveguide allowed to improve the optoelectronic performances of the NIR 2D bP-based photodetector, obtaining a dark current which is lower by three orders of magnitude comparing to the equivalent device based on graphene measured in the same conditions,[182,183] a large bandwidth and a high responsivity.[184]

Thanks to its intrinsic optical anisotropy (*vide supra*), 2D bP can also be used for polarization-sensitive photodetection,[185-188] in a vertical p-n junction device configuration, in a very large spectral range spanning from the visible to the MIR.

2D bP can be efficiently employed also as a broadband light absorbing material in solar cells. In particular, it has been coupled to other semiconducting materials in van der Waals heterostructures to form a heterojunction for excitonic solar cells. In these devices, however, the interfacial band alignment is rather difficult to obtain, being bP/$MoS_2$,[186-190] bP/GaAs [189] and black-red phosphorus heterostructures[190] the unique few examples reported. A promising theoretical work has addressed the photovoltaic properties of a device comprised by a phosphorene monolayer/$TiO_2$ heterostructure leading to a promising power conversion efficiency (PCE).[191,192] In a recent work, 2D bP has been also applied in three-components excitonic solar cells. The integration of liquid exfoliated 2D bP in a Single Wall Carbon Nanotubes (SWCNT)/Si heterojunction led to a strong improvement of the PCE of the corresponding excitonic solar cell (up to 9.37%), thanks to enhanced exciton formation and charge transport and suppressed charge recombination.[193]

The anisotropic dispersion of the complex refractive index in the visible range of 2D bP allowed its application also as photonic component. A theoretical work has analyzedthe optical transmission properties of a photonic structures composed of 2D bP/indium-tin oxide alternated layers.[194] The work demonstrates the emerging of a resonant photonic mode, which envisions the possibility to fabricate of phosphorene-based photonic architectures applicable in sensors and light-emitting devices.

Similarly, 2D bP-based photonic components made of 2D bP sandwiched between two dielectric layers could be employed as switchable optical components. It was demonstrated [195] that in a $SiO_2$/2D bP/$SiO_2$ heterostructure the polariton modes of 2D bP hybridize with those of $SiO_2$. Since the excitation frequency controls the density of photo-induced charges, it is possible to switch the phonon-plasma-polariton mode of the heterostructure, showing potential for its use in high performance ultrafast nanophotonic devices.

In addition, the nonlinear optical (NLO) response of phosphorene to a normally incident, linearly polarized coherent laser radiation, was theoretically demonstrated, with the generation of radiation at tripled frequency accompanied by a specific polarization dependence.[196] Recently, experimental work on a graphene/phosphorene nano-heterojunction was reported, with the aim of combining the ultrafast relaxation,[197] broadband and NLO response of these materials in a 2D nano-optical saturable absorber. The system integrated into an erbium-doped fibre laser generated ultrashort (down to 148 fs) and stable pulses with a ultrahigh energy.[193]

## 9.3. Photocatalytic bP-based systems

As illustrated in the previous paragraphs, 2D bP has a catalytic "van der Waals surface" which can be exploited for a variety of catalytic reactions. In addition to this, the optoelectronic properties allow 2D bP to be used also as efficient photocatalyst, photoelectrocatalyst and photosensitizer. In particular, quantum confinement increases the reducing and oxidizing power of photogenerated carriers in 2D bP with respect to bulk bP. This allowed to demonstrate, in a pioneering work, a 40-fold enhancement in the photocatalysis for the solar-to-chemical energy conversion when transitioning from white phosphorus to 2D bP, in a reaction involving the oxidation of triethylamine (TEA).[50] In the cited work, it was also demonstrated that edge and step defects promote recombination, suggesting a strategy of employing flakes with larger lateral sizes or with chemically passivated edge defects in order to enhance the photocatalytic efficiency. Indeed, a successive theoretical work predicted a highly efficient photocatalytic water splitting by edge pseudohalogen (CN and OCN)-modified phosphorene nanoribbons.[198] It was demonstrated also that the functionalization allowed to tune, through the edge electric dipole, the valence and conduction band-edge energy positions of the system for the efficient photocatalysis of water splitting oxidation and reduction reactions. 2D bP has also been experimentally tested in heterostructures for photocatalysis. In a very recent and groundbreaking work, a bP/BiVO$_4$ 2D heterostructure was designed and applied as a new artificial Z-scheme photocatalytic system able to split water (into H$_2$ and O$_2$ on bP and BiVO$_4$, respectively) under irradiation of visible light, with no need for sacrificial agents or application of an external voltage.[199] (Figure 15)

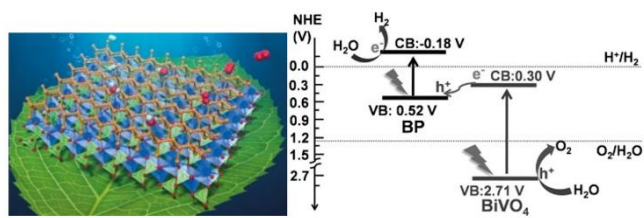

Figure 15. Left: illustration of a water splitting system. Right: Z-scheme photocatalytic water splitting using 2D bP/BiVO$_4$ under visible light irradiation. Reprinted with permission and adapted from Inside Cover of Angew. Chem. Int. Ed. 8/2018 DOI 10.1002/anie.201800579 and M. Zhu, Z. Sun, M. Fujitsuka, T. Majima *Angew. Chem. Int. Ed.* **2018**, *57*, 2160 DOI:10.1002/anie.201711357. Copyright 2018 Wiley.

The photosensitizing properties of 2D bP, that is its singlet oxygen photogeneration ability, combined with its robust biocompatibility,[59,200] allowed to apply this material also for photodynamic therapy (PDT) applications (*vide supra,* Section 5), as photoinducer and binder for radical oxygen species (ROS) for *in situ* disinfection.[201] In detail, an antimicrobial polymeric composite film filled with few-layer bP nanosheets was prepared. It was demonstrated the photosensitization of singlet oxygen at the few-layer bP surface under visible light irradiation and its efficient storage and subsequent thermal release. The system exhibited then a high and rapid disinfection ability in vitro, with antibacterial rate above 99.3% against two different kind of bacteria.

## 10. Conclusions and Perspectives

In this microreview, we have briefly summarized several key aspects of our ongoing research dealing with the chemistry and the application of bulk and exfoliated black phosphorus including an overview of synthesis, electronic structure, high pressure behaviour, decoration with metal nanoparticles and encapsulation in polymers. Some studies addressing the electrical, optoelectronic and catalytic properties of both bP and phosphorene are also described and commented.

Phosphorene may represent a great opportunity for (inorganic) chemists to enter in the two dimensional world as leading actors. From the body of information that we and others have accumulated so far on this material, we strongly believe that an all phosphorus surface has the potential to be functionalized not only by metal nanoparticles, as already shown, but also by metal fragments. This route could drive new exciting possibilities up to the control through a rationale modification of the surface, of the physical properties of the material.

In conclusion, as it is evident from Figure 2, phosphorene has already changed the role of black phosphorus amongst the different allotropes of this element. Instead of the expected ugly duckling, a wonderful black swan has been discovered.

## Acknowledgements

The European Research Council (ERC) and the National Research Council of Italy (CNR) are acknowledged for funding the work through the project PHOSFUN, an ERC Advanced Grant assigned to MP as PI (Grant Agreement No. 670173).

**Keywords:** phosphorene • black phosphorus • 2D material • phosphorus